\def\year{2020}\relax
\documentclass[letterpaper]{article} 
\usepackage{aaai20}  
\usepackage{times}  
\usepackage{helvet} 
\usepackage{courier}  
\usepackage[hyphens]{url}  
\usepackage{graphicx} 
\usepackage{graphicx}  
\usepackage{amsmath}
\usepackage{amsthm}
\usepackage{comment}
\title{Repeated Multimarket Contact with Private Monitoring: A Belief-Free Approach\thanks{
    A full version 
    can be found at http://arxiv.org/abs/1607.03583. 
    Supported in part by KAKENHI 17H01787, 16KK0003, and 17H00761.}}
\author{Atsushi Iwasaki$^{*,{\S}}$, Tadashi Sekiguchi$^{\dagger,{\S}}$, Shun Yamamoto$^{\ddagger}$, and Makoto Yokoo$^{\ddagger,{\S}}$\\
${*}$: University of Electro-Communications, a2c.iwasaki@gmail.com \\
${\dagger}$: Kyoto University, sekiguchi@kier.kyoto-u.ac.jp\\
${\ddagger}$: Kyushu University, \{syamamoto@agent., yokoo@\}inf.kyushu-u.ac.jp \\
${\S}$: RIKEN AIP 
}

\newtheorem{definition}{Definition}

\newtheorem{theorem}{Theorem}
\newtheorem{lemma}{Lemma}

\newtheorem{proposition}{Proposition}

\begin{document}

\maketitle

\begin{abstract}
This paper studies repeated games where two players play multiple duopolistic games simultaneously (multimarket contact). A key assumption is that each player receives a noisy and private signal about the other's actions (\textit{private monitoring} or \textit{observation errors}). 
There has been no game-theoretic support that multimarket contact facilitates collusion or not, in the sense that more collusive equilibria in terms of \textit{per-market} profits exist than those under a benchmark case of one market. 
An equilibrium candidate under the benchmark case is \textit{belief-free} strategies. 
We are the first to construct a non-trivial class of strategies that exhibits the effect of multimarket contact from the perspectives of \textit{simplicity} and \textit{mild punishment}. 
Strategies must be simple because firms in a cartel must coordinate each other with no communication. 
Punishment must be mild to an extent that it does not hurt even the minimum required profits in the cartel. 
We thus focus on two-state automaton strategies such that the players are cooperative in at least one market even when he or she punishes a traitor. 
Furthermore, we identify an additional condition (\textit{partial indifference}), under which the collusive equilibrium yields the optimal payoff.
\end{abstract}

\section{Introduction}
\label{sec:introuction}
This paper investigates a simple but fundamental question: 
Can two players cooperate better when they confront in multiple repeated games
than in a single repeated game?
A typical context we 
concentrate on is \textit{multimarket contact}. 
For example, global enterprises, such as Uber or Lyft, provide their services in multiple distinct markets. 
In each area, they face an oligopolistic competition, which is often modeled as a prisoners' dilemma (PD).
When they repeatedly confront in a long run (\textit{repeated games}), more cooperative or collusive behavior can be an equilibrium.
They may be more likely to collude, and a regulatory agency wants to estimate the extent~\cite{chellappaw:isr:2010}. 

However, the answer to the above question is negative if we assume that each player can directly observe his opponent’s actions (\textit{perfect} monitoring), which is an assumption often used in computer science literature.
\citeauthor{bernheim:rand:1990}~\shortcite{bernheim:rand:1990}
show that multimarket contact does not improve the most collusive
per-market equilibrium profit under perfect monitoring, though some of
the vast empirical studies rather suggest the opposite~\cite{evans:qje:1994}.
To resolve this discrepancy, this paper assumes 
\textit{private} monitoring where each player may observe a different signal. 
For example, although a firm cannot directly observe its rival's action, e.g., prices, 
it can observe a noisy private signal, e.g., its own sales amounts%

Analytical studies on this class of games have not been very successful~\cite{mailath:2006}. 
This is because characterizing all equilibria or identifying optimal
equilibria in games with private monitoring is extremely hard.
Indeed, equilibrium candidates may be so complicated as to be
represented only by automaton strategies with a very large state
space, and the complexity, together with privacy of the other player's
signals, would require very involved statistical inferences to
estimate the other player's history at any period and to check optimality of the continuation strategy~\cite{kandori:2010}.
Notably, a \textit{belief-free} approach has successfully established a general characterization
where an equilibrium strategy is constructed 
so that the statistical inferences do not matter~\cite{ely:jet:2002,ely:econo:2005}. 
However, it is not obvious whether the belief-free approach is helpful in examining
the effects of multimarket contact.
This is because we want to deal with any number of markets, which causes the number of available actions to exponentially increase and may diminish its tractability.

The main goal of this paper is to construct, under multimarket contact with private monitoring, 
a non-trivial class of strategies which can sustain a better per-market outcome than 
an equilibrium strategy for a single market. The secondary concern is twofold. 
First, strategies must be simple. 
Firms do not communicate with each other after they form a cartel 
so that their cartel could be escaped from monitoring by a regulatory agency. 
If they employ a complicated strategy, it will be hard to detect one's deviation and 
to restore its collusion after punishment. 
We thus concentrate on simple two-state automaton strategies, which are still difficult to analyze. 
Second, we assume firms punish a traitor in only some market, instead of all markets. 
When a cartel forms, products therein may not be so differentiated and 
the profit from the market may be small. If a firm punishes a traitor 
in all markets, it may not be able to earn the minimum necessary profit. 
The more collusive equilibrium thus prescribes the players to be
cooperative in at least one market even under a punishment state. 
Surprisingly, our analysis reveals that information from such markets is crucial 
to admit the multimarket contact effect.  

When we employ the belief-free approach, the strategy found by the 
work of Ely and V\"{a}lim\"{a}ki is an important benchmark and it 
attains the optimal payoff among belief-free equilibria in PD~\cite{ely:jet:2002}.
This strategy, which we call EV, can form a belief-free equilibrium 
under private monitoring in the single market case and attains high expected payoffs 
with a wide range of parameter settings. 
Figure~\ref{fig:EV} illustrates EV, which is a variant of the well-known tit-for-tat strategy.%
\footnote{Here, $g$ and $b$ are noisy private signals suggesting that the opponent's action is $C$ and $D$, respectively.
$\varepsilon_R$ or $\varepsilon_P$ represents the transition probability between states. We omit the transitions for the remaining probabilities.} 
%
A player first cooperates and keeps cooperation as long as she observes a signal suggesting cooperation. 
Once she observes a signal suggesting defection, she defects with a given probability and cooperates with the remaining probability.
Similarly, when she defects, she keeps defection as long as she observes a signal suggesting defection.
Once she observes a signal suggesting cooperation, she returns to cooperation with another given probability and defects with the remaining probability.

Building upon those considerations, we provide a condition under which a 
generalization of the benchmark belief-free strategies,
which we call the \textit{generalized EV} (gEV) strategy, forms an equilibrium. 
To the best of our knowledge, we are the first to identify an equilibrium designed for multiple markets 
whose per-market equilibrium payoffs exceed one for the benchmark 
strategies. 
Furthermore, we find an additional condition, which we call \textit{partial indifference}. 
It implies that each player is indifferent among all strategies which differ only in play 
in a given subset of the markets. Under this condition, the gEV equilibrium yields the optimal payoff. 

Let us finally note a simple extension of the EV strategy to multimarket contact. We show that for any number of markets, it is an equally collusive equilibrium under the same condition as the benchmark case. The equilibrium satisfies a special case of partial indifference  (\textit{total indifference}), 
where the subset consists of all markets and is most collusive among such equilibria. Moreover, we reveal that, under the condition, each player's continuation play depends only on the number of signals suggesting the other player's defection in the previous period in a linear manner. This result implies that it must be designed in such a non-linear manner as the gEV strategy. 

\section{Model}
\label{sec:model}
Two players play $M$ PDs simultaneously in each period.
In each PD, each player chooses either $C$ (cooperation) or $D$ (defection). 
This is regarded as a model of oligopolistic competition, where
$C$ is an action increasing the total payoffs 
(for instance, in the case of price competition, charging a collusive high price), and
$D$ is a non-cooperative one (like a price cut).
The players can choose different actions over the $M$ PDs,
so that each player's action set in each period is $\{ C, D \}^M$.

Each player cannot directly observe the other player's actions,
but receives an imperfect signal about them.
In each PD, each player receives either a good signal $g$ or a bad signal $b$.
We assume that each player receives his signals individually, and
cannot observe the other player's signals (private monitoring).
The pair of signals they privately receive in each PD is stochastic, 
following a common symmetric probability distribution that depends entirely on the action pair of that PD.
We denote it by
$o(\omega_1 , \omega_2 |a_1 , a_2 )$, where $(\omega_1 , \omega_2 ) \in \{g, b \}^2$ and $(a_1 , a_2 )\in \{ C, D\}^2$.
We assume that the signals across the $M$ PDs are independent, though
the signals of a given PD may be correlated across the players.  
We also assume that the signal distributions are described by one parameter. 
There exists $p \in (1/2, 1)$ such that for any $i$, any $\omega_j$ ($j \neq i$) and any $a\in \{ C, D\}^2$, 
{\small
\begin{equation*}
\sum_{\omega_i \in \{ g, b\}}o(\omega_i , \omega_j |a )=
\begin{cases}
p & \text{if $(a_i , \omega_j ) \in \big\{ (C, g), (D, b) \big\}$,}\\
1-p & \text{otherwise.}
\end{cases}
\end{equation*}}
The marginal distribution of an individual signal in a given PD is such that the \textit{right} signal ($\omega_j
=g$ if $a_i =C$, and $\omega_j =b$ if $a_i =D$) is received with probability $p$.  We let $s =1-p$, which is the
probability of an \textit{error}.
The assumption is consistent with \textit{conditionally independent} monitoring, 
which is a representative monitoring structure.
Formally, a signal distribution is conditionally independent 
if $o(\omega_i, \omega_j\mid a) = o(\omega_i\mid a)o(\omega_j\mid a)$ 
for all $\omega_i$, $\omega_j$, and $a$. 
Also, it is consistent with nearly perfect monitoring (when $p$ is close to $1$),
but inconsistent with nearly public monitoring (namely, the case where the event $\omega_1 =\omega_2$ is much more
likely than $\omega_1 \neq \omega_2$). 

In each PD, player~$i$'s payoff depends only on his action and the signal of that PD. 
The payoff function is common to all PDs, denoted by $\pi_i (a_i , \omega_i )$.
We are more interested in the expected payoff function: 
\[\abovedisplayskip=2pt\belowdisplayskip=2pt
g_i (a_1, a_2)=\sum_{(\omega_1 , \omega_2 )}\pi_i (a_i , \omega_i )o(\omega_1 , \omega_2 |a_1 , a_2 ).
\]
We assume that their expected payoff functions are represented by the following payoff matrix: 
\begin{center}
    \begin{tabular}[htbp]{|c|c|c|}\hline
        & $C$ & $D$ \\\hline
    $C$ & $1,1$ & $-y,1+x$ \\\hline
    $D$ & $1+x$,$-y$ & $0,0$ \\\hline
    \end{tabular}
\end{center}
\noindent We assume $x>0$, $y>0$ and $1>x-y$, so that it indeed represents a PD. 

All $M$ PDs are played infinitely, in periods $t=0,1,2,\ldots$.  Player~$i$'s {\it private history} at the
beginning of period $t \ge 1$ is an element of $H_{i}^t \equiv \big[ \{ C, D \}^M \times \{g, b\}^M \big]^{t}$. Let
$H_{i}^0$ be an arbitrary singleton, and let $H_i =\cup_{t \ge 0}H_{i}^t$ be the set of player~$i$'s all private histories.  
Player~$i$'s strategy of this repeated game is a mapping from $H_i$ to the set of all probability distributions over $\{ C, D \}^M$.
That is, we allow randomized strategies. 
If the actual play of the repeated game is such that
the action pair $\big( a_{1}^m (t), a_{2}^m (t) \big)$ is played in the $m$-th PD in period~$t$ for each $m$ and $t$, 
player~$i$'s normalized average payoff is 
\begin{equation}\label{eq: repeated game path payoff} \abovedisplayskip=2pt\belowdisplayskip=2pt
(1-\delta )\sum_{t=0}^{\infty}\delta^t \sum_{m=1}^M g_i \big( a_{1}^m (t), a_{2}^m (t) \big) ,
\end{equation}
where $\delta \in (0,1)$ is their common discount factor. The average payoff of any strategy pair is the expected
value of Eq.~\ref{eq: repeated game path payoff}, where the expectation is taken with respect to the players'
randomizations and the monitoring structure.

\subsection{Belief-Free Equilibrium}\label{sec:belief-free}

The solution concept for repeated games with imperfect monitoring is
\textit{sequential equilibrium}~\cite{RePEc:ecm:emetrp:v:50:y:1982:i:4:p:863-94}.
However, since it is still highly difficult to analyze our model, 
we here focus on a special class called {\it belief-free equilibria}~\cite{ely:econo:2005},
which is standard in the private monitoring literature~\cite{mailath:2006}. 

\begin{definition}[Belief-Free Equilibrium]
  A strategy pair is a {\it belief-free equilibrium} if for any $t \ge 0$, $h_{1}^t \in H_{1}^t$ and $h_{2}^t \in H_{2}^t$, 
  each player~$i$'s continuation strategy given $h_{i}^t$ is optimal against player~$j$'s continuation strategy given $h_{j}^t$.  
\end{definition}

An important property 
is that, 
while player~$i$ given her private history should, in principle, 
optimize her continuation payoff against her belief about player~$j$'s history (and hence his continuation strategy), 
her continuation strategy is optimal even if she were to know $j$'s history with certainty.%
\footnote{We refer to player~$i$ or 1 as \textit{her} and to player~$j$ or 2 as \textit{him} throughout this paper.} 
In other words, the players playing a belief-free equilibrium need not compute their beliefs in the course of play. 
When a strategy pair is represented by finite-state automaton strategies, as will be the case in subsequent analysis, 
it is a belief-free equilibrium if any player's continuation strategy
(behavior expanded from the automaton) starting from any state
is a best response (optimal) against the other player's continuation strategy starting from any state. 
Note that we never restrict the other's possible strategy space, which includes strategies with an infinite number of states. 

Suppose both players employ a common strategy represented by a two-state automaton with state space $\{ R, P\}$.
Let $V_{s_1 s_2}$, where $s_1 \in \{ R, P\}$ and $s_2 \in \{ R, P\}$, be player~1's continuation payoff when
(i) player~2 is currently at $s_2$ and then follows the automaton, and
(ii) player~1 always plays the action prescribed at state~$s_1$ at any subsequent history.
The strategy pair is a belief-free equilibrium if and only if there exist $V_R$ and $V_P$ such that 
\begin{equation}\label{eq: payoffs of constant strategies} 
V_{RR}=V_{PR}=V_R , \quad V_{RP}=V_{PP}=V_P ,
\end{equation}
and that $V_{s_2}$ ($s_2 \in \{ R, P\}$) is player~1's best response payoff against
player~2's continuation strategy when he is at state~$s_2$. 
To see this, note that by Eq.~\ref{eq: payoffs of constant strategies}, player~1 at any history is indifferent
between her continuation strategy at state~$R$ and that at state~$P$ irrespective of her belief about player~2's state.
Since the second condition implies that both continuation strategies give her best response payoff at any history,
the conditions for belief-free equilibrium are all satisfied.

We shall consider a general class of two-state automaton strategies throughout this paper.
\begin{definition}[Two-State Automaton Strategies]
  \label{def:FSA}
The state space is $\{ R, P\}$, and $R$ is the initial state.
At state~$s\in\{R,P\}$, the player is prescribed to choose $a^s\in\{C,D\}^M$.
Suppose the current state is $R(P)$. If $\omega=(\omega_k)_{k=1}^M\in\{g,b\}^M$ is observed, then the
state shifts to $P(R)$ with probability $\xi_R(\omega)$ ($\xi_P(\omega)$) 
(and stays at $R(P)$ with the remaining probability).
Note that this class is parameterized by $\xi_R(\cdot)$ and $\xi_P(\cdot)$: For any $s\in\{R,P\}$, $\xi_s:\{g,b\}^M\rightarrow [0,1]$,
which we call the transition probability functions. 
\end{definition}

\begin{figure}[tb]
  \centering
  \begin{minipage}[htbp]{0.41\linewidth}
  \centering
  \includegraphics[width=0.88\linewidth]{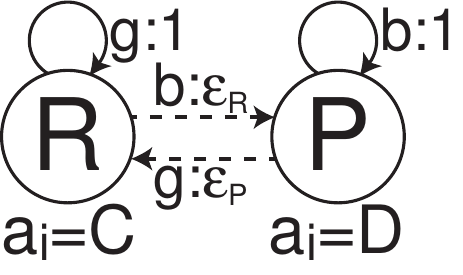}   
 \vspace{-1.0mm}
  \caption{
    EV strategy} 
  \label{fig:EV}
  \end{minipage}
  \begin{minipage}[htbp]{0.49\linewidth}
  \centering
  \includegraphics[width=0.86\linewidth]{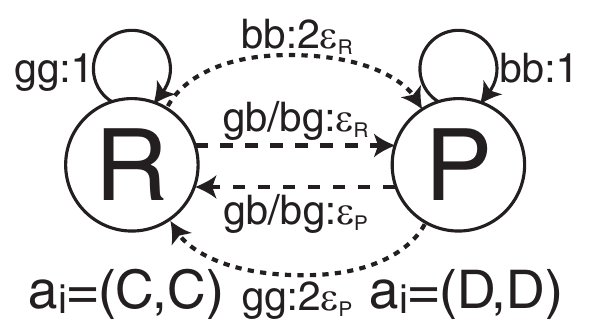}
   \vspace{-1.5mm}
  \caption{sEV strategy}
  \label{fig:sEV}
  \end{minipage}
\end{figure}

\subsection{Ely-V\"{a}lim\"{a}ki Strategy} 
\label{sec:ev}

Let us next explain an instance of this class for a single PD case ($M=1$). Figure~\ref{fig:EV} illustrates the EV strategy~\cite{ely:jet:2002}
where $a^R=C$, $a^P=D$, $\xi_R(g)=1$, $\xi_R(b)=\varepsilon_R$, $\xi_P(b)=1$, and $\xi_P(g)=\varepsilon_P$.
Note that
\begin{align*}
  & \varepsilon_R = \frac{(1-\delta)x}{\delta\Big\{ 2p-1-(1-p)(x+y) \Big\}}, \text{and} \\
  & \varepsilon_P = \frac{(1-\delta)y}{\delta\Big\{ 2p-1-(1-p)(x+y) \Big\}}. 
\end{align*}
A solid line denotes a deterministic transition and a dashed line denotes a probabilistic transition, 
though, for simplicity, we omit some state transitions. 
EV is a representative two-state automaton strategy that forms a belief-free equilibrium 
under repeated games with private monitoring and attains the highest average payoff among belief-free equilibria in PD. 
A player first cooperates at state $R$, but after observing a bad signal, 
she punishes (defects) at the next period with probability $\varepsilon_{R}$, or keep cooperation with $1-\varepsilon_{R}$. 
Likewise, after she defects at $P$, 
if she observes a good signal, she returns cooperation with $\varepsilon_{P}$, 
or keep defection with $1-\epsilon_{P}$.
%
\begin{proposition}\label{prop:EV}
There exist $\varepsilon_{R}\in [0,1]$ and $\varepsilon_{P}\in [0,1]$ such that 
the EV strategy pair is a belief-free equilibrium if 
\begin{equation}
  \label{eq:EV-condition}
  \delta \big[ 2p-1 - (1-p)(x+y) +\max \{ x, y \} \big] \geq  \max \{ x, y \}. 
\end{equation}
The average payoff starting from state $R$ is 
\begin{equation*}
V_R = V^{EV} \equiv 1 - \frac{(1-p)x}{2p-1}. 
\end{equation*}
\end{proposition}

\section{Simplified EV Equilibrium}

What happens if there are $M(\geq 2)$ PDs, in comparison with the case of one PD? 
If EV forms an equilibrium, it is always an equilibrium to play it in each PD independently. 
Obviously, the payoff of this equilibrium is $M$ times the EV equilibrium payoff.
Under this equilibrium, a player's actions in all PDs can be quite different,
depending on the histories of individual PDs. Thus, the corresponding automaton has $2^M$ states.

Interestingly, this equilibrium strategy can be greatly simplified 
so that it is an equilibrium with the same payoff and under the same condition. 
The simplified strategy just has two states, where 
a player cooperates in all PDs at one state and defects in all PDs at the other. 
Her actions are therefore perfectly correlated across the PDs. 
The class of strategies, which we call \textit{simplified EV} (sEV), is simple, 
also in the sense that its transition probabilities from one state to the other 
depend only on the number of bad or good signals. 
Therefore, the player need not to know the exact configuration of the signals. 

\begin{definition}[sEV Strategy]
  An sEV strategy for $M(\geq 2)$ PDs is a two-state automaton strategy
  parameterized by two numbers $\varepsilon_{R}\in [0,1]$ and $\varepsilon_{P}\in [0,1]$: 
  \begin{itemize}
  \item The actions at each state are prescribed by $a^R=(C,C,\ldots,C)$ and $a^P=(D,D,\ldots,D)$, 
  \item The transition probabilities are defined as
    \begin{align*}
      \xi_R(\omega)=|\{k\mid \omega_k=b\}|\varepsilon_R, \text{and } \xi_P(\omega)=|\{k\mid \omega_k=g\}|\varepsilon_P.
    \end{align*}
\end{itemize}
\end{definition}
Figure~\ref{fig:sEV} illustrates sEV for two PDs in the same manner as Figure~\ref{fig:EV}. 
The next theorem identifies the equilibrium condition and the average payoff starting from state $R$. 
\begin{theorem}
  \label{thm:sEV}
  There exist $\varepsilon_{R}\in [0,1]$ and $\varepsilon_{P}\in [0,1]$ such that 
  the sEV strategy pair is a belief-free equilibrium if Eq.~\ref{eq:EV-condition} holds. 
  The average payoff starting from state $R$ is $M V^{EV}$
\end{theorem}
We place the proof in the full version 
because it is similar to Proposition~\ref{prop:EV}. 

\section{Generalized EV Equilibrium and \\ Partially Indifference}

The goal of the analysis of this section is to give one answer to whether a class of strategies which
can yield a better per-market payoff than the EV strategy exists or not. 
Let us define the class of strategies as the \textit{generalized EV} (gEV) strategy
that achieves the optimal payoff 
{\small\begin{align}\label{eq:overlineYd}
    \overline{Y}^d \equiv M - \frac{1-p}{2p-1}(d+1)x.
\end{align}}
We also use 
{\small\begin{align}\label{eq:underlineYd}
    \underline{Y}^d \equiv M-d + \frac{1-p}{2p-1}dy + \frac{p^{M-d}}{p^{M-d}-(1-p)^{M-d}}(M-d)x.
\end{align}}
\begin{definition}[gEV Strategy]\label{def:gEV}
  A gEV strategy for $M(\geq 2)$ PDs is a two-state automaton strategy parameterized by five parameters, $d\in\{1,2,\ldots,M-1\}$, 
  $\alpha^R$, $\alpha^P$, $\beta^R$, and $\beta^P$: 
  {\small
  \begin{align*}
    & \alpha^{R} = \frac{(1-\delta)x}{\delta (\overline{Y}^d-\underline{Y}^d)(2p-1)}, \\
    & \alpha^{P} = \frac{(1-\delta)y}{\delta (\overline{Y}^d-\underline{Y}^d)(2p-1)}, \\
    & \beta^{R} = \frac{(1-\delta)x}{\delta (\overline{Y}^d-\underline{Y}^d)(2p-1)(1-p)^{M-d-1}}, \\
    & \beta^{P} = \frac{(1-\delta)(M-d)x}{\delta (\overline{Y}^d-\underline{Y}^d)\Big\{p^{M-d}-(1-p)^{M-d}\Big\}}.
\end{align*}}
  Let us divide the markets into $A = \{1,2,\ldots, M-d\}$ and $B=\{M-d+1,M-d+2,\ldots,M\}$.
  The actions at each state are prescribed by $a^R=(\overbrace{C,C,\ldots,C}^M)$ and $a^P=(\overbrace{C,\ldots,C}^{M-d},\overbrace{D,\ldots,D}^d)$.
  The transition probabilities are defined as
   {\small
  \begin{align*}
    \xi_R(\omega) =
    \begin{cases}
      \alpha^R|\{k \in B \mid \omega_k=b \}| + \beta^R 
      \, \text{if  $\omega_{k}=b$ for all $k\in A$}, \\
      \alpha^R|\{k\in B \mid \omega_k=b \}| \quad\quad\quad \text{otherwise}. 
    \end{cases}
  \end{align*}
  \begin{align*}
    \xi_P(\omega) =
    \begin{cases}
      \alpha^P|\{k \in B \mid \omega_k=g \}| + \beta^P 
      \, \text{if $\omega_{k}=g$ for all $k\in A$}, \\
      \alpha^P|\{k \in B \mid \omega_k=g \}| \quad\quad\quad \text{otherwise}. 
    \end{cases}
  \end{align*} }
\end{definition}

Let us explain how we construct this strategy. 
A player cooperates in $A$ at state $P$. Then, she always cooperates PDs in $A$ regardless of which state she is in. 
The transition probabilities from $R$ to $P$ distinguish signals from $A$ with from $B$. 
The increase of transition probabilities is constant for the number of bad signals from $A$. 
If she observes at least one good signal from $A$, it is zero, otherwise, $\beta^{R}$. 
The transition probabilities further increase by $\alpha^R$ in the number of bad signals from $B$. 
Similarly, the transition probabilities from $P$ to $R$ are specified. 
Their increase is constant for the number of good signals from $A$. 
If she observes at least one bad signal from $A$, it is zero, otherwise, $\beta^P$. 
The transition probabilities increase by $\alpha^P$ in the number of good signals from~$B$.
 
The following theorem identifies a condition for an equilibrium
outperforming the sEV equilibrium. 
\begin{theorem}\label{thm:gEV-existence}
  If $\overline{Y}^d > \underline{Y}^d$, 
  there exists $\underline{\delta}\in (0,1)$ such that for any $\delta\geq \underline{\delta}$,
  there is a 
  belief-free equilibrium with $V_R=\overline{Y}^d$ and $V_P=\underline{Y}^d$.
  %
\end{theorem}
The payoffs equal the upper and the lower bounds derived in Theorems~\ref{thm:optimal payoff} and \ref{thm:worst payoff}. Those bounds of equilibrium payoffs are obtained 
under the following condition, which we call \textit{partial indifference}. 
Recall that $A=\{1,2,\ldots,M-d\}$ and $B=\{M-d+1,M-d+2,\ldots,M\}$ by parameter $d$. 
Let $V_s(a)$ be the continuation payoff when player 2 is at state~$s$ and player 1 chooses action $a$. 
\begin{definition}[Partial Indifference]\label{def:PIBF}
  Fix $d\in [1,M-1]$.
  An equilibrium is \textit{$d$-partially indifferent} if 
  for any $s\in\{R,P\}$, any $a',a''\in\{C,D\}^M$, and any $k \in A$, if $a'_k=a''_k$, $V_s(a')=V_s(a'')$. 
\end{definition}

The $d$-partial indifference means that, given the other player's strategy, a player's repeated game payoff does not depend on the play of the $B$ markets once he fixes a play of the $A$ markets. For example, if the player cooperates in all $A$ markets (as in the equilibrium), he receives a common repeated game payoff regardless of his actions in the $B$ markets. Or if he defects in all $A$ markets, he receives a common repeated game payoff regardless of his actions in the $B$ markets. Similar indifference holds for any play of the $A$ markets. In other words, this concept implies that his repeated game payoff given the other player's strategy depends solely on his play in the $A$ markets. Further, our analysis shows that the repeated game payoff is maximized when the player cooperates in all $A$ markets. 
Thus, the strategy is optimal among all possible strategies. 
\mdseries

Let us outline the proof of Theorem~\ref{thm:gEV-existence}. 
We calculate both the upper (lower) bound of the continuation payoffs 
from state $R$ ($P$) 
among all $d$-partially indifferent equilibria among the class of strategies in Definition~\ref{def:FSA} (Theorems~\ref{thm:optimal payoff} and \ref{thm:worst payoff}). 
Since our equilibrium uses a transition to $R$ as a reward and a transition to $P$ as a punishment, the upper bound must exceed the lower bound (Theorem~\ref{thm:VR-VP positive}). Under that condition, we construct a particular gEV equilibrium which achieves those upper and lower bounds 
if the discount factor is sufficiently large.

Suppose that $a^R$ and $a^P$
are prescribed as in Definition~\ref{def:gEV}.
Formally, given $d$, $a^R=(a^R_k)_{k=1}^M$, $a^P=(a^P_k)_{k=1}^M$ 
where for all $k$, $a_k^R=C$, for all $k\in A$, $a_k^P=C$, and for all $k\in B$, $a_k^P=D$. 
\begin{theorem}\label{thm:optimal payoff}
  Fix $d\in [1,M-1]$. 
  Suppose $a^R$ is prescribed as in Definition~\ref{def:gEV}. 
  Any $d$-partially indifferent belief-free equilibrium payoff is at most $\overline{Y}^d$ defined in Eq.~\ref{eq:overlineYd}. 
\end{theorem}

\begin{proof}
  Consider an arbitrary $d$-partially indifferent belief-free equilibrium. Fix $k\in B$. 
  For any $a_{-k}\in\{C,D\}^{M-1}$, $d$-partially indifferent belief-freeness implies 
  {\small
  \begin{align}\label{eq:belief-free for k at R}
    (1-\delta)x = \delta (V_R-V_P) \Big\{ \zeta_R(D,a_{-k}) - \zeta_R(C,a_{-k}) \Big\}. 
  \end{align}}
  
  Note that $o_2(\omega_{-k} \mid a_{-k} )$ is the probability that player~2 observes a signal profile $\omega_{-k}$ from the
  markets except $k$ occurs when player 1 chooses an action profile $a_{-k}$. 
  Besides $\zeta_s(a)$ indicates 
  $\sum_{\omega\in\{g,b\}^M} o_2(\omega \mid a)\xi_s(\omega)$ for state $s$. 
  Then we derive for any $a_{-k}$, 
  {\small
  \begin{align}\label{eq:bf_at_R}
  (1-\delta)x &= \delta (V_R-V_P)(2p-1)\sum_{\omega_{-k}\in\{g,b\}^{M-1}}\Big\{ \xi_R(b,\omega_{-k}) \\\notag &- \xi_R(g,\omega_{-k}) \Big\} o_2(\omega_{-k}\mid a_{-k}).  
  \end{align}}
  
  We have, for any $a_{-k}\in \{C,D\}^{M-1}$, $\zeta_R(D,a_{-k}) - \zeta_R(C,a_{-k})=$
  {\small
  \begin{align*}
    (2p-1)\sum_{\omega_{-k}\in\{g,b\}^{M-1}}\Big\{ &\xi_R(b,\omega_{-k}) - \\ &\xi_R(g,\omega_{-k}) \Big\}o_2(\omega_{-k}\mid a_{-k}). 
  \end{align*}}
  Substituting this into Eq.~\ref{eq:belief-free for k at R}, we obtain 
  \begin{align*}
    \xi_R(b,\omega_{-k}) - \xi_R(g,\omega_{-k}) = \frac{(1-\delta)x}{\delta (V_R-V_P)(2p-1)}. 
  \end{align*}
  Intuitively, this implies the increase of the transition probabilities is constant when
  the signal observed in a market changes from the good to the bad one.
  Since we can arbitrarily choose $k\in B$, 
  there exists a function $\beta^R: \{g,b\}^{M-d}\rightarrow [0,1]$ such that
  {\small\begin{align*}
    \xi_R(\omega) = \frac{(1-\delta)x|\{k\in B \mid\omega_k=b\}|}{\delta(V_R-V_P)(2p-1)} + \beta^R (\omega_1,\ldots,\omega_{M-d})
  \end{align*}} for all $\omega \in \{g,b\}^M$. Note that the transition probability from state $R$ to $P$ is linear in the number of bad signals from~$B$. 

  Let us consider the incentive condition that a player at state $R$ defects in a single market. 
  Recall that $a^R=(C,\ldots,C)$ and let $a'^R$ be an action that defects only in a market and cooperates in the remaining, e.g., $(D,C,\ldots,C)$. 
  \begin{align*}
    V_R& \geq (1-\delta)(M+x) + \delta V_R - \delta (V_R-V_P)\zeta_R(a'^R) \\ 
    \leftrightarrow & (1-\delta)x \leq \delta (V_R-V_P) \Big\{ \zeta_R(a'^R) - \zeta_R(a^R) \Big\}. 
  \end{align*}
  Since the left hand side of this equation is positive, we have $\zeta_R(a'^R) > \zeta_R(a^R)$. 
  For $a \in \{a^R, a'^R\}$, it holds that
  \[
    \zeta_R(a) = \frac{(1-\delta)x(1-p)d}{\delta (V_R-V_P)(2p-1)} + \eta_R(a) 
  \] where $\eta_R(a) = $ 
  {\small
  \[
  \sum_{(\omega_1,\ldots,\omega_{M-d})\in \{g,b\}^{M-d}} \left[ \prod_{k=1}^{M-d} o_2(\omega_k\mid a)\right] \beta^R(\omega_1,\ldots,\omega_{M-d}).
  \]}
  Clearly, $\zeta_R(a'^R)-\zeta(a^R)=\eta_R(a'^R)-\eta_R(a^R)>0$ holds.
  From those equations, let us transform the expected payoff starting from state $R$. 
  \begin{align*}
    V_R = & M - \frac{\delta}{1-\delta}(V_R-V_P)\zeta_R(a^R) \\ 
    \leq  & M - \frac{1-p}{2p-1}xd - \frac{x}{\eta_R(a'^R)-\eta_R(a^R)}\eta_R(a^R).
  \end{align*}
  Note that $\eta_R(a)$ must be non-zero. If not, for any $(\omega_1,\ldots,\omega_{M-d})\in \{g,b\}^{M-d}$,
  $\beta^R(\omega_1,\ldots, \omega_{M-d})=0$ and for any two $a,a'\in \{C,D\}^{M}$, $\eta(a')-\eta(a)=0$. 
  This contradicts $\eta_R(a'^R)-\eta_R(a^R)>0$. 
  Therefore, there exists a signal profile $(\omega_1,\ldots,\omega_{M-d})\in\{g,b\}^{M-d}$ such that $\beta^R(\omega_1,\ldots,\omega_{M-d})>0$.
  We have
  \begin{align*}
    1 < \frac{\eta_R(a'^R)}{\eta_R(a^R)} 
    \leq \frac{p}{1-p}. 
  \end{align*}
  Accordingly, we finally obtain $V_R \leq \overline{Y}^d$. 
  The proof is complete. 
\end{proof}

The upper bound always exceeds the benchmark, unless $d=M-1$. 
In this equilibrium, the fact that the $d$-partial indifference decreases 
the continuation payoff from $R$ 
by $\frac{1-p}{2p-1}dx$ follows from the linearity of the transition probabilities in $B$.
The remaining term of $\frac{1-p}{2p-1}x$ follows from the incentive condition that a player chooses $C$ in $A$.

Let us next derive the lower bound. 
\begin{theorem}\label{thm:worst payoff}
  Fix $d\in [1,M-1]$.
  Suppose $a^P$ is prescribed as in Definition~\ref{def:gEV}. 
  Among the class of strategies in Definition~\ref{def:FSA}, 
  any $d$-partially indifferent belief-free equilibrium payoff is at least $\underline{Y}^d$
  defined in Eq.~\ref{eq:underlineYd}. 
\end{theorem}
Since the flow of this proof is the same as that of Theorem~\ref{thm:optimal payoff},
we place it in the full version. 

Before proving Theorem~\ref{thm:gEV-existence}, 
we show in any $d$-partially indifferent belief-free equilibrium, 
$\overline{Y}^d \geq V_R > V_P \geq \underline{Y}^d$ must hold.
Otherwise, none of the equilibrium exists. To this end, it is suffice to 
the following statement holds. 
\begin{theorem}\label{thm:VR-VP positive}
  Fix $d\in [1,M-1]$. 
  In any $d$-partially indifferent belief-free equilibrium payoff, $V_R >V_P$. 
\end{theorem}

\begin{proof}
  Consider an arbitrary $d$-partially indifferent belief-free equilibrium.
  Recall that $\zeta_s(a)$ is $\sum_{\omega\in\{g,b\}^M} o_2(\omega \mid a)\xi_s(\omega)$ for state $s$. 
  From Definition~\ref{def:FSA}, we have
  \begin{align*}
    &V_R=(1-\delta)M + \delta V_R - \delta (V_R-V_P) \zeta_R(a^R),\ \textrm{and}\\
    &V_P=(1-\delta)(M-d) + \delta V_P + \delta (V_R-V_P) \zeta_P(a^P) 
  \end{align*}
  hold and we obtain 
  \begin{align*}
    V_R- V_P = \frac{(1-\delta)d}{(1-\delta) +\delta \Big\{ \zeta_R(a^R)+\zeta_P(a^P) \Big\} }. 
  \end{align*}
  Thus, $V_R>V_P$ holds because the right hand side is clearly positive. 
\end{proof}

\begin{proof}[Proof of Theorem~\ref{thm:gEV-existence}]
  Fix $d\in [1,M-1]$ and suppose the gEV strategy. 
  We claim that $\overline{Y}^d>\underline{Y}^d$ implies $\xi_s(\omega)\in [0,1]$ for all $s\in \{R,P\}$ and $\omega\in \{g,b\}^M$. 
  It is immediate that $\xi_s(\cdot)$ is always positive if $\overline{Y}^d>\underline{Y}^d$. 

  Next, define $\underline{\delta}\in (0,1)$ as $\delta$ satisfying 
  \begin{align*}
    \max \Big\{ \alpha^Rd+\beta^R, \alpha^Pd+\beta^P \Big\} = 1. 
  \end{align*}
  Since $d$ is the maximum value of $|\{k\in B \mid \omega_k=b \}|$ or $|\{k\in B \mid \omega_k=g \}|$, for any $\delta\geq \underline{\delta}$,
  it holds that $\xi_s(\omega) \leq 1$ for all $s\in \{R,P\}$ and $\omega\in \{g,b\}^M$. 
  Therefore, the gEV strategy is verified.

  By solving the following system of the equations, we obtain $V_R=\overline{Y}^d$ and $V_P=\underline{Y}^d$. 
  \begin{align*}
    V_R & = (1-\delta) g_R(a^R) + \delta V_R -\delta(V_R-V_P)\zeta_R(a^R) \\
        & = (1-\delta) g_R(a^P) + \delta V_R -\delta(V_R-V_P)\zeta_R(a^P). \\
    V_P & = (1-\delta) g_P(a^R) + \delta V_P + \delta(V_R-V_P)\zeta_P(a^R) \\
        & = (1-\delta) g_P(a^P) + \delta V_P + \delta(V_R-V_P)\zeta_P(a^P). 
  \end{align*}

  It suffices to verify that
  (i) $V_R$ is a player's best response payoff when the other player is at state $R$, and
  (ii) $V_P$ is a player's best response payoff when the other player is at state $P$.
  To this end, let $V_s(d_A , d_B)$ be a player's continuation payoff when
  he chooses $D$ at some $d_A$ PDs in $A$ 
  at some $d_B$ PDs in $B$. 
  Then it conforms to the strategy from the next period on, given that the other player is at state~$s$ and conforms to the strategy.
  The proof is complete if we show that $V_s(0, 0)\ge V_s(d_A , d_B)$ for any $d_A$, any $d_B$, and any $s$.
  
  First, some calculation verifies
  {\small
  \begin{align*}
    V_R(d_A, d_B)= 
         (1-\delta )M + \delta V_R - \delta(V_R-V_P)d(1-p)\alpha^R \\
      +  (1-\delta)x \Big[ d_A - \Big(\frac{p}{1-p}\Big)^{d_A}\frac{1-p}{2p-1}\Big]. 
   \end{align*}}
 Hence, $V_R(d_A, d_B)$ does not depend on $d_B$. For any $d_A$ and $d_B$, $V_R(d_A, d_B) - V_R(d_A+1, d_B) =$
 {\small
 \begin{align*}
    (1-\delta)x\Big[ -1 + \Big( \frac{p}{1-p} \Big)^{d_A} \Big]. 
 \end{align*}}
  Since $p > 1-p$, $V_R(d_A, d_B) \geq V_R(d_A+1, d_B)$ holds for any $d_A$ and $d_B$. 
  We obtain $V_R(d_A,d_B) = $ 
  {\small
  \begin{align*}
    & \sum_{k=1}^{d_A} \Big\{ V_R(k,d_B) - V_R(k-1,d_B) \Big\} + V_R(0, d_B) 
                 \leq V_R(0,d_B) 
  \end{align*}}
  for any $d_A$ and $d_B$, as desired. 
  Further, it attains the same value $-(1-\delta)x\frac{1-p}{2p-1}$ at $d_A=0$ and $d_A=1$.
  The concavity of the last term of $V_R(d_A, d_B)$ implies this consequence.

  Finally, some calculation verifies 
  {\small
  \begin{align*}
     V_P&(d_A,d_B) = \\ 
      & (1-\delta )(M-d-dy) + \delta V_P - \delta(V_R-V_P)dp\alpha^P \\
    + & (1-\delta)x \Big[ d_A + \Big( \frac{1-p}{p} \Big)^{d_A} \frac{p^{M-d}(M-d)}{p^{M-d}-(1-p)^{M-d}} \Big]. 
  \end{align*}}
Hence, $V_P(d_A, d_B)$ does not depend on $d_B$. 
For any $d_B$, it attains the same value
\[(1-\delta)x\frac{p^{M-d}(M-d)}{p^{M-d}-(1-p)^{M-d}}\]
at $d_A=0$ and $d_A=M-d$. 
Since $p>1-p$, it is convex in $d_A$. This convexity implies that $V_P(0,d_B) \geq V_P(d_A,d_B)$ for any $d_A$ and $d_B$.
The proof is complete.
\end{proof}

Theorem~\ref{thm:gEV-existence} reaches the main goal we raised at the beginning of this paper.
Namely, if there exists an integer $d<M-1$ such that $\overline{Y}^d>\underline{Y}^d$, the gEV strategy yields
the average per-market payoff greater than the EV and sEV strategies. Note that the number of markets $M$ must be
greater than three. 
The condition $\overline{Y}^d>\underline{Y}^d$ is equivalent to
\begin{align*} 
  d\Big\{ 1- \frac{1-p}{2p-1}(x+y)\Big\} 
           & > \Big\{ \frac{1-p}{2p-1}  \\\notag  +& (M-d)\frac{p^{M-d}}{p^{M-d}-(1-p)^{M-d}}\Big\}x .
\end{align*}
The coefficient of $d$ in the left hand side must be positive because that of $x$ in the right hand side is positive.
This is the necessary condition that an EV equilibrium exists for some discount factor from Eq.~\ref{eq:EV-condition}. 
Unless EV is an equilibrium, for any $d$, no $d$-partially indifferent belief-free equilibrium exists.

\subsection{sEV Equilibrium and Total Indifference}

This subsection turns to the sEV equilibrium to deepen the understandings of gEV. 
Consider an extreme case of partial indifference. We restrict an equilibrium such that every action is
indifferent in the continuation payoff regardless of the opponent's state. 
\begin{definition}[Total Indifference]\label{def:TIBF}
  An equilibrium is \textit{totally indifferent} if for any $s\in\{R,P\}$, any $a',a''\in\{C,D\}^M$, $V_s(a')=V_s(a'')$. 
\end{definition}
%
sEV 
clearly is totally indifferent. In addition, the next theorem claims the bound of the equilibrium payoff. 
\begin{theorem}
  \label{prop:TIBF payoff}
  Suppose $a^R=(C,C,\ldots,C)$. Among the class of strategies in Definition~\ref{def:FSA}, 
  any totally indifferent belief-free equilibrium payoff is at most $M V^{EV}$. 
\end{theorem}
We place the proof in the full version. 
We emphasize that the total indifference requires the signal in one
market to have the same impact on the transition probabilities as the
signal in any other market, as is seen from
\[ 
\xi_R(\omega)= \frac{(1-\delta)x|\{k\mid\omega_k=b\}|}{\delta(V_R-V_P)(2p-1)} + \beta^R. 
\]
Hence, the sEV
equilibrium is most collusive among all totally indifferent equilibria because
it has $\beta^R =0$ 
and therefore is least likely to switch to inefficient punishment.
Since Theorem~\ref{prop:TIBF payoff} does not specify behavior in state $P$, mild punishment is not sufficient to improve the per-market equilibrium payoff. Some nonlinearity of the transition probabilities is necessary. Signals from markets, where each player is prescribed to cooperate at any state, are essential to admit more collusive equilibria.

\section{Discussions}
\label{sec:discussions}

Let us explain why gEV can form an equilibrium.
A key feature involves the nonlinearity of the transition probabilities. 
In fact, the transition probabilities from state~$R$ to $P$ do not depend on the outcome in $A$ at all, 
as long as it contains at least one good signal. 
However, if all signals from $A$ are bad, the transition probability sharply increases by $\beta^R$. 
Why does this nonlinearity help? 
Suppose the other player is at state~$R$, and consider how a player wants to play the PDs in $A$. 
Her incentive to play $C$ or $D$ in one PD in $A$ crucially depends on the probability of the event that 
all signals among the other PDs in $A$ are bad. 
Only under that event, is her action in this PD pivotal. 
Naturally, the event is more likely when she defects among more PDs in $A$.
Therefore, her temptation to defect in one PD in $A$ is largest when she cooperates among all other PDs in $A$.
Note that we apply a similar argument to this when we check the incentives. 
This observation implies that once gEV prevents a player from defecting in one PD in $A$, 
it automatically ensures that the player has no incentive to defect in any number of PDs in $A$. 
Thus, as long as we consider gEV, 
we can effectively ignore all actions which defect among two or more PDs in $A$. 
This reduction in the number of incentive constraints is a key to the payoff improvement results brought about by gEV.

Next let us note why our gEV strategy outperforms the sEV strategy. 
First of all, if we had zero A market (d=M), gEV would be equal to sEV. 
Thus, gEV with a few A markets is a natural equilibrium candidate. 
Further, perpetual cooperation in the $A$ markets leads to an improvement of the per-market equilibrium payoff.  
Relatedly, the presence of the $A$ markets is a key part of our results and reflects some reality. A typical example is airline industries: cutting the fares in every route rarely occurs (except some promotion or campaign), and the fares in routes connecting major cities tend to be almost the same across airlines~\cite{evans:qje:1994}.
\mdseries

Another question is whether gEV achieves an optimal payoff among much more general equilibrium strategies than $d$-partially indifferent equilibria.
Under perfect 
monitoring, it is known that a player's equilibrium payoff vector can be computed. 
Dynamic programming can derive the bounds of the equilibrium payoff of each player, i.e., a self-generation set~\cite{abreu:econometrica:1990}.
The existence of an equilibrium strategy is guaranteed that attains a payoff vector in that set. 
Unfortunately, under private monitoring, this is generally impossible because the recursive structure
under perfect or public monitoring does not persist. Checking the optimality is the immediate future work. 

\subsection{Related literature}
In the literature of computer science, AI, and multi-agent systems, 
there are many streams associated with repeated games~\cite{burkov:ker:2013}: 
The complexity of equilibrium computation~\cite{littman:dss:2005,borgs:geb:2010,andersen:aaai:2013}, 
multi-agent learning~\cite{blum:agt:2007,conitzer:ml:2007,shoham::2008}, 
partially observable stochastic games (POSGs)~\cite{hansen:aaai:2004,%
  doshi:aaai:2006,tennenholtz:AAMAS:2009,mescheder:bcai:2011,wunder:aamas:2011}, and so on. 
Among them, POSGs are most relevant because repeated games with private monitoring can be considered as a special case of POSGs. 
However, POSGs often impose partial observability on an opponent's strategy (behavior rule) 
and not on opponent's past actions~\cite{mescheder:bcai:2011,wunder:aamas:2011}. 
They estimate an optimal (best reply) strategy against an unknown strategy (not always fixed) 
from perfectly observable actions (perfect monitoring). 
In contrast, we verify whether a given strategy profile is a \textit{mutual} best reply after any history, 
i.e., finding an equilibrium, with partially observable actions (private monitoring). 
Thus, this paper also addresses understanding the gap between POSGs and 
repeated games with private monitoring in economics. 
  
In fact, very few existing works have addressed verifying an equilibrium. 
\citeauthor{hansen:aaai:2004}~\shortcite{hansen:aaai:2004} 
develop an algorithm that iteratively eliminates dominated strategies. 
However, just eliminating dominated strategies is not sufficient to
find an equilibrium. Also, the algorithm is not applicable to
an infinitely repeated game.
\citeauthor{doshi:aaai:2006}~\shortcite{doshi:aaai:2006} 
investigate the computational complexity of achieving equilibria in interactive POMDPs. 

The economics literature has extensively studied another interesting
class of repeated games where the players observe a common noisy
signal (\textit{public} monitoring).  \citeauthor{kobayashi:geb:2012}~\shortcite{kobayashi:geb:2012} studied this version of our model and showed that when
multimarket contact facilitates collusion, the most collusive
equilibrium payoff is attained by a variant of the trigger strategy.
Hence, the players must defect in all markets in the punishment state.
Our model has an opposite implication which rather favors mild punishment.

Alternatively, another important topic has been validity of the folk theorem.  
Most of them assume perfect or public monitoring, see a textbook \cite{mailath:2006}. 
In case of private monitoring, \citeauthor{sugaya:mimeo:2015}~\shortcite{sugaya:mimeo:2015} establishes a general folk theorem.
However, the result is irrelevant to our analysis because the equilibrium strategies are excessively
complicated and require nearly complete patience of the players. 
Specializing in multimarket contact, we rather show that the gEV strategy forms
a highly cooperative equilibrium and only requires the players to be mildly patient.

\section{Conclusions}

This paper examined equilibria in multimarket contact with a noisy signal. 
To the best of our knowledge, under private monitoring, we are the first to find the multimarket contact effect, i.e.,
the existence of more collusive equilibria than the single market case. 
We constructed the gEV strategy and clarified the structure of the equilibria, by finding the partial indifferent condition, 
which leads that strategy to the best possible payoff. 
In future works, we are particularly interested in an extension to asymmetric markets. 
We believe our equilibrium construction easily extends to the case of asymmetric markets, 
only at the expense of additional notations. 
Under asymmetry, the colluding firms may want to optimally choose the $A$ markets and the $B$ markets, 
which makes our problem more complicated.


\bibliographystyle{aaai}
\bibliography{multimarket,jaws2014-yamamoto,monitoring}

\begin{thebibliography}{}

\bibitem[\protect\citeauthoryear{Abreu, Pearce, and
  Stacchetti}{1990}]{abreu:econometrica:1990}
Abreu, D.; Pearce, D.; and Stacchetti, E.
\newblock 1990.
\newblock Toward a theory of discounted repeated games with imperfect
  monitoring.
\newblock {\em Econometrica} 58:1041--1064.

\bibitem[\protect\citeauthoryear{Andersen and
  Conitzer}{2013}]{andersen:aaai:2013}
Andersen, G., and Conitzer, V.
\newblock 2013.
\newblock Fast equilibrium computation for infinitely repeated games.
\newblock In {\em Proceedings of the AAAI Conference on Artificial Intelligence
  (AAAI)},  53--59.

\bibitem[\protect\citeauthoryear{Bernheim and
  Whinston}{1990}]{bernheim:rand:1990}
Bernheim, B.~D., and Whinston, M.~D.
\newblock 1990.
\newblock {Multimarket Contact and Collusive Behavior }.
\newblock {\em RAND Journal of Economics} 21(1):1--26.

\bibitem[\protect\citeauthoryear{Blum and Monsour}{2007}]{blum:agt:2007}
Blum, A., and Monsour, Y.
\newblock 2007.
\newblock Learning, regret minimization, and equilibria.
\newblock In {\em Algorithmic game theory}. Cambridge University Press.
\newblock  79--101.

\bibitem[\protect\citeauthoryear{Borgs \bgroup et al\mbox.\egroup
  }{2010}]{borgs:geb:2010}
Borgs, C.; Chayes, J.; Immorlica, N.; Kalai, A.~T.; Mirrokni, V.; and
  Papadimitriou, C.
\newblock 2010.
\newblock The myth of the folk theorem.
\newblock {\em Games and Economic Behavior} 70(1):34 -- 43.

\bibitem[\protect\citeauthoryear{Burkov and Chaib-draa}{2013}]{burkov:ker:2013}
Burkov, A., and Chaib-draa, B.
\newblock 2013.
\newblock Repeated games for multiagent systems: A survey.
\newblock {\em The Knowledge Engineering Review}  1--30.

\bibitem[\protect\citeauthoryear{Chellappaw, Sambamurthy, and
  Saraf}{2010}]{chellappaw:isr:2010}
Chellappaw, R.~K.; Sambamurthy, V.; and Saraf, N.
\newblock 2010.
\newblock Competing in crowded markets: Multimarket contact and the nature of
  competition in the enterprise systems software industry.
\newblock {\em Information Systems Research} 21(3):614--630.

\bibitem[\protect\citeauthoryear{Conitzer and
  Sandholm}{2007}]{conitzer:ml:2007}
Conitzer, V., and Sandholm, T.
\newblock 2007.
\newblock {AWESOME: a general multiagent learning algorithm that converges in
  self-play and learns a best response against stationary opponents}.
\newblock {\em Machine Learning} 67(1):23--43.

\bibitem[\protect\citeauthoryear{Doshi and
  Gmytrasiewicz}{2006}]{doshi:aaai:2006}
Doshi, P., and Gmytrasiewicz, P.~J.
\newblock 2006.
\newblock {On the Difficulty of Achieving Equilibrium in Interactive {POMDPs}}.
\newblock In {\em Proceedings of the 21st National Conference on Artificial
  Intelligence (AAAI)},  1131--1136.

\bibitem[\protect\citeauthoryear{Ely and V\"{a}lim\"{a}ki}{2002}]{ely:jet:2002}
Ely, J.~C., and V\"{a}lim\"{a}ki, J.
\newblock 2002.
\newblock A robust folk theorem for the prisoner's dilemma.
\newblock {\em Journal of Economic Theory} 102(1):84--105.

\bibitem[\protect\citeauthoryear{Ely, Horner, and
  Olszewski}{2005}]{ely:econo:2005}
Ely, J.~C.; Horner, J.; and Olszewski, W.
\newblock 2005.
\newblock Belief-free equilibria in repeated games.
\newblock {\em Econometrica} 73(2):377--415.

\bibitem[\protect\citeauthoryear{Evans and Kessides}{1994}]{evans:qje:1994}
Evans, W.~N., and Kessides, I.~N.
\newblock 1994.
\newblock {Living by the "Golden Rule": Multimarket Contact in the U.S. Airline
  Industry}.
\newblock {\em The Quarterly Journal of Economics} 109(2):341--366.

\bibitem[\protect\citeauthoryear{Hansen, Bernstein, and
  Zilberstein}{2004}]{hansen:aaai:2004}
Hansen, E.~A.; Bernstein, D.~S.; and Zilberstein, S.
\newblock 2004.
\newblock Dynamic programming for partially observable stochastic games.
\newblock In {\em Proceedings of the 19th National Conference on Artificial
  Intelligence (AAAI)},  709--715.

\bibitem[\protect\citeauthoryear{Kandori}{2010}]{kandori:2010}
Kandori, M.
\newblock 2010.
\newblock Repeated games.
\newblock In Durlauf, S.~N., and Blume, L.~E., eds., {\em Game theory}.
  {Palgrave Macmillan}.
\newblock  286--299.

\bibitem[\protect\citeauthoryear{Kobayashi and Ohta}{2012}]{kobayashi:geb:2012}
Kobayashi, H., and Ohta, K.
\newblock 2012.
\newblock {Optimal collusion under imperfect monitoring in multimarket
  contact}.
\newblock {\em Games and Economic Behavior} 76(2):636--647.

\bibitem[\protect\citeauthoryear{Kreps and
  Wilson}{1982}]{RePEc:ecm:emetrp:v:50:y:1982:i:4:p:863-94}
Kreps, D.~M., and Wilson, R.
\newblock 1982.
\newblock Sequential equilibria.
\newblock {\em Econometrica} 50(4):863--94.

\bibitem[\protect\citeauthoryear{Littman and Stone}{2005}]{littman:dss:2005}
Littman, M.~L., and Stone, P.
\newblock 2005.
\newblock {A polynomial-time Nash equilibrium algorithm for repeated games}.
\newblock {\em Decision Support Systems} 39(1):55--66.

\bibitem[\protect\citeauthoryear{Mailath and Samuelson}{2006}]{mailath:2006}
Mailath, G., and Samuelson, L.
\newblock 2006.
\newblock {\em Repeated Games and Reputation}.
\newblock Oxford University Press.

\bibitem[\protect\citeauthoryear{Mescheder, Tuyls, and
  Kaisers}{2011}]{mescheder:bcai:2011}
Mescheder, D.; Tuyls, K.; and Kaisers, M.
\newblock 2011.
\newblock {POMDP} opponent models for best response behavior.
\newblock In {\em Proceedings of the 23rd Benelux Conference on Artificial
  Intelligence}.

\bibitem[\protect\citeauthoryear{Shoham and Leyton-Brown}{2008}]{shoham::2008}
Shoham, Y., and Leyton-Brown, K.
\newblock 2008.
\newblock Learning and teaching.
\newblock In {\em Multiagent systems: Algorithmic, Game-Theoretic, and Logical
  Foundations}. Cambridge University Press.
\newblock  189--222.

\bibitem[\protect\citeauthoryear{Sugaya}{2015}]{sugaya:mimeo:2015}
Sugaya, T.
\newblock 2015.
\newblock Folk theorem in repeated games with private monitoring.
\newblock Revised and resubmitted to Review of Economic Studies.

\bibitem[\protect\citeauthoryear{Tennenholtz and
  Zohar}{2009}]{tennenholtz:AAMAS:2009}
Tennenholtz, M., and Zohar, A.
\newblock 2009.
\newblock Learning equilibria in repeated congestion games.
\newblock In {\em Proceedings of the 8th International Joint Conference on
  Autonomous Agents and Multi-Agent System (AAMAS)},  233--240.

\bibitem[\protect\citeauthoryear{Wunder \bgroup et al\mbox.\egroup
  }{2011}]{wunder:aamas:2011}
Wunder, M.; Kaisers, M.; Yaros, J.~R.; and Littman, M.
\newblock 2011.
\newblock Using iterated reasoning to predict opponent strategies.
\newblock In {\em The 10th International Conference on Autonomous Agents and
  Multiagent Systems (AAMAS)},  593--600.

\end{thebibliography}

\newpage

\appendix
\section{Proof of Theorem~\ref{thm:sEV}}

\begin{proof}
Suppose Eq.~\ref{eq:EV-condition} holds. Let us define 
\begin{gather*}
V_R =M\bigg\{ 1-\frac{(1-p)x}{2p-1} \bigg\} , \quad V_P =M\frac{(1-p)y}{2p-1}, \\
\varepsilon_{R}=\frac{(1-\delta )x}{\delta (2p-1)(V_R -V_P )}, \quad 
\varepsilon_{P}=\frac{(1-\delta )y}{\delta (2p-1)(V_R -V_P )} .
\end{gather*}
From Eq.~\ref{eq:EV-condition}, we obtain 
\begin{align*}
V_R >V_P , \quad 0<M\varepsilon_R \le 1, \quad 0 <M\varepsilon_P \le 1.
\end{align*}
Hence, the sEV strategy with $\varepsilon_R$ and $\varepsilon_P$ above is well-defined.

From the definition of sEV, some calculations verify that
{\small
\abovedisplayskip=3.0pt\belowdisplayskip=3.0pt
\begin{align*}
V_{RR} & =  (1-\delta )M  +\delta V_{RR} -\delta M(1-p) \varepsilon_R (V_{RR}-V_{RP}), \\
V_{RP} & = -(1-\delta )My +\delta V_{RP} +\delta Mp\varepsilon_P (V_{RR}-V_{RP}), \\
V_{PR} & =(1-\delta )(1+x)M +\delta V_{PR} -\delta Mp \varepsilon_R (V_{PR}-V_{PP}), \\
V_{PP} & = \delta V_{PP} +\delta M (1-p)\varepsilon_P (V_{PR}-V_{PP}).
\end{align*}}
Solving these, we obtain Eq.~\ref{eq: payoffs of constant strategies}. 

Let $V_s(d)$ be a player's payoff when he defects in $d$ PDs and then conforms to sEV,
given that the other player's current state is $s \in \{ R, P\}$.
The proof is complete if we show that $V_s \ge V_s(d)$ for any $s$ and any $d$. 
It is easy to verify that 
 {\small 
\begin{align*}
     V_R(d) & =(1-\delta )(M +dx) +\delta V_R \\ & -\delta \big\{ dp +(M-d)(1-p) \big\} \varepsilon_R (V_R -V_P ).
\end{align*}}
When we regard it as a linear function of $d$, its slope is 
\begin{equation*}
(1-\delta )x -\delta (2p-1) \varepsilon_R (V_R -V_P )=0, 
\end{equation*}
where the equality follows from the definition of $\varepsilon_R$. Therefore, $V_R(d) =V_R$ for any $d$, as desired. 
Similarly, it is easily seen that 
 {\small
\begin{align*} 
V_P(d) &=-(1-\delta )(M -d)y +\delta V_P  \\ &+\delta \big\{ (M-d)p+d(1-p) \big\} \varepsilon_P (V_R -V_P ). 
\end{align*} }
Its slope as a function of $d$ is 
\begin{equation*}
(1-\delta )y -\delta (2p-1) \varepsilon_P (V_R -V_P )=0, 
\end{equation*}
where the equality follows from the definition of $\varepsilon_P$.
Therefore, $V_P(d) =V_P$ for any $d$, which completes the proof. 
\end{proof}

\section{Proof of Theorem~\ref{thm:worst payoff}}

\begin{proof}
  Fix $d\in [1,M-1]$ and $k \in B$. 
  Consider an arbitrary $d$-partially indifferent belief-free equilibrium. 
  For any $a_{-k}\in\{C,D\}^{M-1}$, $d$-partially indifferent belief-freeness implies 
  \begin{align}
    \label{eq:belief-free for k at P}
                        (1-\delta)y = \delta (V_R-V_P) \Big\{ \zeta_P(C,a_{-k}) - \zeta_P(D,a_{-k}) \Big\}. 
  \end{align}
  From 
  Eq.~\ref{eq:bf_at_R}, since we have, for any $a_{-k}\in \{C,D\}^{M-1}$, $\zeta_P(C,a_{-k}) - \zeta_P(D,a_{-k}) = $ 
  \begin{align*}
    (2p-1)\sum_{\omega_{-k}\in\{g,b\}^{M-1}}\Big\{&\xi_P(g,\omega_{-k}) \\ 
    - &\xi_P(b,\omega_{-k}) \Big\} o_2(\omega_{-k}\mid a_{-k}).
  \end{align*}
  Substituting this into Eq.~\ref{eq:belief-free for k at P}, we obtain 
  \begin{align*}
    \xi_P(g,\omega_{-k}) - \xi_P(b,\omega_{-k}) = \frac{(1-\delta)y}{\delta (V_R-V_P)(2p-1)}. 
  \end{align*}
  Intuitively, this implies the increase of the transition probabilities is constant when
  the signal observed in a market changes from the bad to the good one. 

  Since we can arbitrarily choose $k$, which is greater than $M-d$,
  there exists a function $\beta^P: \{g,b\}^{M-d}\rightarrow [0,1]$ such that
  for all $\omega \in \{g,b\}^M$,
  {\small\begin{align*}
    \xi_P(\omega) = \frac{(1-\delta)y|\{k\in B\mid\omega_k=g\}|}{\delta(V_R-V_P)(2p-1)} + \beta^P (\omega_1,\ldots,\omega_{M-d})
  \end{align*}}
  holds.
  Note that the transition probability from state $R$ to $P$ is linear with respect to the number of bad signals observed in $B$

  Let us consider the incentive condition that a player at state $P$ defects in all the markets.
  Suppose that $a^P$ is an action that cooperates in $A$ and defects in $B$, i.e., $a^P=(\overbrace{C,\ldots,C}^{M-d}, \overbrace{D,\ldots, D}^d)$,
  and that $a'^P$ is one that defects in all the markets, i.e., $a'^P=(\overbrace{D,\ldots,D}^M)$. 
  \begin{align*}
    V_P = & (1-\delta)(M-d) + \delta V_P + \delta (V_R-V_P)\zeta_P(a^P) \\
    \geq & (1-\delta)(M-d)(1+x) + \delta V_P + \delta (V_R-V_P)\zeta_P(a'^P) \\ 
    \leftrightarrow & (1-\delta)(M-d)x \leq \delta (V_R-V_P) \Big\{ \zeta_P(a^P) - \zeta_P(a'^P) \Big\}
  \end{align*}
  Since the left hand side of this equation is positive, we have $\zeta_P(a^P) > \zeta_P(a'^P)$.
  For $a \in \{a^P, a'^P\}$, it holds that
  \[
    \zeta_P(a) = \frac{(1-\delta)y(1-p)d}{\delta (V_R-V_P)(2p-1)} + \eta_P(a) 
  \] where $\eta_P(a) = $ 
  {\small
  \[
    \sum_{(\omega_1,\ldots,\omega_{M-d})\in \{g,b\}^{M-d}} \left[ \prod_{k=1}^{M-d} o_2(\omega_k\mid a)\right] \beta^P(\omega_1,\ldots,\omega_{M-d}).
  \]}
  Clearly, $\zeta_P(a^P)-\zeta(a'^P)=\eta_P(a^P)-\eta_P(a'^P)>0$ holds.
  Utilizing those equations, let us transform the expected payoff starting from state $P$.
  \begin{align*}
    V_P = & M-d + \frac{\delta}{1-\delta}(V_R-V_P)\zeta_P(a^P) \\ 
        = & M-d + \frac{1-p}{2p-1}yd + \frac{\delta}{1-\delta}(V_R-V_P)\eta_P(a^P) \\
    \geq & M-d + \frac{1-p}{2p-1}yd + \frac{(M-d)x}{\zeta_P(a^P)-\zeta_P(a'^P)}\eta_P(a^P) \\
    =    & M-d + \frac{1-p}{2p-1}yd + \frac{(M-d)x}{\eta_P(a^P)-\eta_P(a'^P)}\eta_P(a^P). 
  \end{align*}

  Note that $\eta_P(a)$ must be non-zero. If not, for any $(\omega_1,\ldots,\omega_{M-d})\in \{g,b\}^{M-d}$,
  $\beta^P(\omega_1,\ldots, \omega_{M-d})=0$ and for any two $a,a'\in \{C,D\}^{M}$, $\eta_P(a)-\eta_P(a')=0$. 
  This contradicts $\eta_P(a^P)-\eta_P(a'^P)>0$.
  Therefore, there exists a signal profile $(\omega_1,\ldots,\omega_{M-d}\in\{g,b\}^{M-d})$
  such that $\beta^P(\omega_1,\ldots,\omega_{M-d})>0$.
  We have
  \begin{align*}
    1 < \frac{\eta_P(a^P)}{\eta_P(a'^P)} 
    \leq  \left( \frac{p}{1-p} \right)^{M-d}. 
  \end{align*}
  Accordingly, we finally obtain 
  {\small
  \begin{align*}
    V_P \geq & M-d + \frac{1-p}{2p-1}yd + (M-d)x \frac{
               \frac{\eta_P(a^P)}{\eta_P(a'^P)}
               }{\frac{\eta_P(a^P)}{\eta_P(a'^P)}-1} \\
    \geq & M-d + \frac{1-p}{2p-1}yd + (M-d)x \frac{
           \left(\frac{p}{1-p}\right)^{M-d}
           }{\left(\frac{p}{1-p}\right)^{M-d}-1} \\
    \geq & M-d + \frac{1-p}{2p-1}yd + (M-d)x \frac{
           p^{M-d}
           }{p^{M-d}-(1-p)^{M-d}}.
         \end{align*}}
  The proof is complete.
\end{proof}

\section{Proof of Theorem~\ref{prop:TIBF payoff}}

Before proving this theorem, let us introduce the following lemma.
\begin{lemma}\label{lem:regularity}
  Let $\pi_{M}$ ($M\geq 1$) be a $2^M\times 2^M$ matrix parameterized by $p\in (1/2,1)$ and 
  it is inductively defined as follows:
  {\small
  \[
    \pi_{1} = 
    \begin{pmatrix}
      p & 1-p \\
      1-p & p
    \end{pmatrix},\ 
    \pi_{M+1} = 
    \begin{pmatrix}
      p\pi_{M} & (1-p)\pi_{M} \\
      (1-p)\pi_{M} & p\pi_{M}
    \end{pmatrix}. 
  \]}
  For each $M$, $\pi_M$ is regular. 
\end{lemma}
We place the proof in the next section. 

\begin{proof}
  Let us consider a totally indifferent belief-free equilibrium and define the followings: 
  Let $V_{s}$ ($s\in\{R,P\}$) be the player 1's continuation payoff that is her best response
  against the player 2's continuation strategy starting from state~$s$. 
  Let $g_s(a)$ ($a\in\{C,D\}^M$) be the stage game (expected) payoff when player 1 and 2 choose $a$ and $a^s$, respectively.
  Let $\zeta_s(a)$ be $\sum_{\omega\in\{g,b\}^M} o_2(\omega \mid a)\xi_s(\omega)$. 
  This is the probability that the state of the player 2 shifts to the opposite when player 1 chooses $a$.
  Note that $o_2(\omega \mid a)$ is the probability that player 2 observes signals $\omega$ from $M$ markets when player 1 chooses $a$. 
  
  Fix $k\in [1,M]$. From Definition~\ref{def:TIBF}, for any $a_{-k}\in\{C,D\}^{M-1}$, which is an action profile except market $k$, 
  {\small
  \begin{align}
    V_R & = (1-\delta) g_R(C, a_{-k}) + \delta V_R -\delta(V_R-V_P)\zeta_R(C,a_{-k}) \label{eq:V_R_C}\\
        & = (1-\delta) g_R(D, a_{-k}) + \delta V_R -\delta(V_R-V_P)\zeta_R(D,a_{-k}) \notag 
  \end{align}}
  We then have 
  {\small
  \begin{align*}
    (1-\delta)x& = \delta (V_R-V_P)\Big\{ \zeta_R(D,a_{-k}) - \zeta_R (C,a_{-k})\Big\} \\
               &= \, \delta (V_R-V_P)\sum_{\omega\in\{g,b\}^M} \Big\{ o_2(\omega \mid D,a_{-k} ) \\
&~~~~~~~~~~~~~~~~~~~~~~~~~~~~~~~~~~~~~~~~~ - o_2(\omega \mid C,a_{-k}) \Big\}. 
  \end{align*}}
  For any $\omega_{-k}\in\{g,b\}^M$,
  \begin{align*}
    & \Big\{ o_2(\omega \mid D,a_{-k} ) - o_2(\omega \mid C,a_{-k}) \Big\}\xi_R(\omega)  \\ & = 
    \begin{cases}
      (2p-1)o_2(\omega_{-k} \mid a_{-k} )\xi_R(b,\omega_{-k}) \mathrm{\ if\ } \omega_k=b, \\
      -(2p-1)o_2(\omega_{-k} \mid a_{-k} )\xi_R(g,\omega_{-k}) \mathrm{\ if\ } \omega_k=g.
    \end{cases}
  \end{align*}

  Eq.~\ref{eq:bf_at_R} holds only when $V_R\neq V_P$ from $p\neq \frac{1}{2}$ and contains a
  $\left( o_2(\omega_{-k}\mid a_{-k}) \right)_{\omega_{-k}\in\{g,b\}^{M-1}}$
  for any $a_{-k}\in\{C,D\}^{M-1}$, which is a $2^{M-1}$-dimensional vector. 
  Further, it can be supposed to have a $2^{M-1}\times 2^{M-1}$ matrix 
    \[
      \left(\left( o_2(\omega_{-k}\mid a_{-k}) \right)_{\omega_{-k}\in\{g,b\}^{M-1}} \right)_{a_{-k}\in\{C,D\}^{M-1}}.
    \]
  Lemma~\ref{lem:regularity} verifies that the matrix is regular. 
  Choosing for any $\omega_{-k}\in \{g,b\}^{M-1}$. 
    \begin{align}
      \label{eq:affinity}
      \xi_R(b,\omega_{-k}) - \xi_R(g, \omega_{-k}) = \frac{(1-\delta)x}{\delta (V_R-V_P)(2p-1)}
    \end{align}
    clearly satisfies Eq.~\ref{eq:bf_at_R}.
  Thus, the linear independence ensures the uniqueness of $2^{M-1}$-dimensional vector of 
    $\left( \xi_R(b,\omega_{-k}) - \xi_R(g, \omega_{-k}) \right)_{\omega_{-k}\in\{g,b\}^{M-1}}$.
  Because Eq.~\ref{eq:affinity} holds for an arbitrary $k$, there exists $\beta^R\geq 0$ such that
    \begin{align}
      \label{eq:3}
      \xi_R(\omega)= \frac{(1-\delta)x|\{k\mid\omega_k=b\}|}{\delta(V_R-V_P)(2p-1)} + \beta^R 
    \end{align}
    holds for all $\omega\in\{g,b\}^M$.
  This is derived by the fact that, given the signal profile $\omega_{-k}$, 
  the increase of the transition probability is constant when the signal from a market changes from good to bad. 
  The transition probability from $R$ to $P$ is guaranteed to be linear in the number of bad signals
  observed in all the markets. 
  
  Suppose that $a_{-k}=(C,\ldots,C)$. Eq.~\ref{eq:V_R_C} is transformed into
  \begin{align*}
    V_R &= (1-\delta)M + \delta V_R -\delta(V_R-V_P)\zeta_R(a^R) \\
        &= M - \frac{\delta}{1-\delta}(V_R-V_P)\zeta_R(a^R) \\
        &= M - \frac{\delta}{1-\delta}(V_R-V_P)\Big( \frac{(1-\delta)x\cdot M(1-p)}{\delta(V_R-V_P)(2p-1)} + \beta^R \Big)\\
        &\leq M \Big(1- \frac{1-p}{2p-1}x\Big) = M V^{EV}
  \end{align*} The proof is complete. \end{proof}

\section{Proof of Lemma~\ref{lem:regularity}}

\begin{proof}
  $\pi_1$ is clearly regular (the determinant is non-zero) because $p\neq 1/2$. 
  Fix $M > 1$. Suppose that $\pi_M$ is regular.
  So, if a profile of weights $(\lambda_k)_{k=1}^{2^M}$ satisfies
  $\sum_{k=1}^{2^M}\lambda_k\pi_M^k=\mathbf{0}$, it must hold that $\lambda_1=\lambda_2=\cdots=\lambda_{2^M}=0$. 
  Note that $\pi_M^k$ is a $2^M$ dimensional vector in the $k$-th component of 
  \[
    \pi_M =
    \begin{pmatrix}
      \pi_M^1 \cdots \pi_M^k \cdots \pi_M^{2^M} \\
    \end{pmatrix}^T. 
  \]
  Next, by induction, we have 
  \[
    \pi_{M+1} =
    \begin{pmatrix}
      p\pi_M^1         & (1-p)\pi_M^1     \\
      \vdots           & \vdots          \\
      p\pi_M^{2^M}      & (1-p)\pi_M^{2^M} \\
      (1-p)\pi_M^1      & p\pi_M^1     \\
      \vdots           & \vdots          \\
      (1-p)\pi_M^{2^M}  & p\pi_M^{2^M} \\
    \end{pmatrix}. 
  \]

  Let us show that, for a profile of $2^{M+1}$ weights $\mathbf{\mu}$ and $\mathbf{\nu}$, 
  \begin{align*}
    (\mu_1,\ldots,\mu_{2^{M+1}}, \nu_1,\ldots,\nu_{2^M}) \pi_{M+1} = \mathbf{0}. 
  \end{align*}
  This equation is decomposed into the two equations with $\pi_{M}$:
  {\small
  \begin{align*}
    p \sum_{k=1}^{2^M} \mu_k \pi_{M}^k + (1-p)\sum_{k=1}^{2^M} \nu_k \pi_{M}^k 
                                     &= \sum_{k=1}^{2^M} \{p\mu_k+(1-p)\nu_k\}\pi_{M}^k = \mathbf{0}, \\
    (1-p)\sum_{k=1}^{2^M} \mu_k \pi_{M}^k + p\sum_{k=1}^{2^M} \nu_k \pi_{M}^k 
                                        &= \sum_{k=1}^{2^M} \{(1-p)\mu_k+p\nu_k\}\pi_{M}^k = \mathbf{0} .
  \end{align*}}
  Since $\pi_M$ is regular, it holds that for any $k$, 
  \begin{align*}
    p\mu_k+(1-p)\nu_k = (1-p)\mu_k+p\nu_k = 0. 
  \end{align*}
  This implies $(2p-1)(\mu_k-\nu_k)=0$ and we have $\mu_k=\nu_k$ for all $k$. 
  We then obtain $\mu_k=\nu_k=0$ for all $k$. 
  Accordingly, $\pi_{M+1}$ is regular. The proof is complete. 
\end{proof}

\end{document}